\documentclass[10pt,conference]{IEEEtran}
\IEEEoverridecommandlockouts

\usepackage[utf8]{inputenc}
\usepackage[cjk]{kotex}
\usepackage{graphicx}		
\usepackage{float}			
\usepackage{subfloat}		
\usepackage{subfigure}		
\usepackage{gensymb}
\usepackage{amsmath}
\usepackage{amssymb}
\usepackage{amsthm}
\usepackage{exscale}
\usepackage{textcomp}		
\usepackage{bm}
\usepackage{algorithmic}
\usepackage[linesnumbered,ruled,noend,vlined]{algorithm2e} 
\usepackage[switch]{lineno}
\usepackage{soul}
\usepackage{ragged2e}
\usepackage{array}
\usepackage{relsize}
\usepackage{multirow}
\usepackage{booktabs}		
\usepackage{siunitx}
\usepackage{url}
\usepackage{cite}
\usepackage[symbol]{footmisc}
\usepackage{lipsum}

\theoremstyle{plain}

\theoremstyle{definition}

\theoremstyle{remark}

\theoremstyle{problem}

\usepackage[textsize=footnotesize]{todonotes}

\usepackage{graphicx}
\usepackage{pgfplots}
\usepackage{tikz}
\usepackage{adjustbox}
\usetikzlibrary{calc}
\usetikzlibrary{positioning}

\def\BibTeX{{\rm B\kern-.05em{\sc i\kern-.025em b}\kern-.08em
    T\kern-.1667em\lower.7ex\hbox{E}\kern-.125emX}}
\begin{document}

\title{QoS-Aware Graph Contrastive Learning for Web Service Recommendation\\
\thanks{Duksan Ryu is the corresponding author. This research was supported by Basic Science Research Program through the National Research Foundation of Korea (NRF) funded by the Ministry of Education (NRF- 2022R1I1A3069233) and the MSIT (Ministry of Science and ICT), Korea, under the ITRC (Information Technology Research Center) support program (IITP-2023-2020-0-01795) supervised by the IITP (Institute for Information \& Communications Technology Planning \& Evaluation) and the Nuclear Safety Research Program through the Korea Foundation Of Nuclear Safety (KoFONS) using the financial resource granted by the Nuclear Safety and Security Commission (NSSC) of the Republic of Korea. (No. 2105030)}
}

\author{
\IEEEauthorblockN{Jeongwhan Choi}
\IEEEauthorblockA{\textit{Dept. of Artificial Intelligence} \\
\textit{Yonsei University}\\
Seoul, South Korea\\
jeongwhan.choi@yonsei.ac.kr}
\and
\IEEEauthorblockN{Duksan Ryu}
\IEEEauthorblockA{\textit{Dept. of Software Engineering} \\
\textit{Jeonbuk National University}\\
Jeonju, South Korea \\
duksan.ryu@jbnu.ac.kr}
}

\maketitle

\begin{abstract}
With the rapid growth of cloud services driven by advancements in web service technology, selecting a high-quality service from a wide range of options has become a complex task. This study aims to address the challenges of data sparsity and the cold-start problem in web service recommendation using Quality of Service (QoS). We propose a novel approach called QoS-aware graph contrastive learning (QAGCL) for web service recommendation. Our model harnesses the power of graph contrastive learning to handle cold-start problems and improve recommendation accuracy effectively. By constructing contextually augmented graphs with geolocation information and randomness, our model provides diverse views. Through the use of graph convolutional networks and graph contrastive learning techniques, we learn user and service embeddings from these augmented graphs. The learned embeddings are then utilized to seamlessly integrate QoS considerations into the recommendation process. Experimental results demonstrate the superiority of our QAGCL model over several existing models, highlighting its effectiveness in addressing data sparsity and the cold-start problem in QoS-aware service recommendations. Our research contributes to the potential for more accurate recommendations in real-world scenarios, even with limited user-service interaction data.
\end{abstract}

\begin{IEEEkeywords}	
Quality of Service, Web Service, Service Recommendation, Graph Contrastive Learning
\end{IEEEkeywords}

\section{Introduction}
Quality of Service (QoS) is a crucial aspect of web service technologies. Web services are reusable web components designed to support machine-to-machine interaction through programmable method calls~\cite{zhangservices}. As reported by \textit{ProgrammableWeb}, there are 20,525 public web services, and their availability is accelerating with the advancement of cloud computing~\cite{duan2012survey}. Many of these web services offer similar functionalities to the users. QoS represents the quality attributes of web services and is perceived as an essential criterion for distinguishing these services. 
Predicting QoS is a very popular and active research area~\cite{ghafouri2022survey,tang2015wswalker,lee2015lmf,ryu2018location,choi2022gain}. Various approaches have been proposed through QoS prediction, such as service recommendation~\cite{zheng2012collaborative}, selection~\cite{yu2007efficient}, and discovery~\cite{phalnikar2012survey}.
As web service technology advances rapidly, the cloud has become a repository of a vast array of service options~\cite{Zhang2010WSExpress}. 
Selecting an appropriate service among various options, mainly based on QoS, poses a significant challenge.

\begin{figure}
  \centering
  \includegraphics[width=0.56\columnwidth]{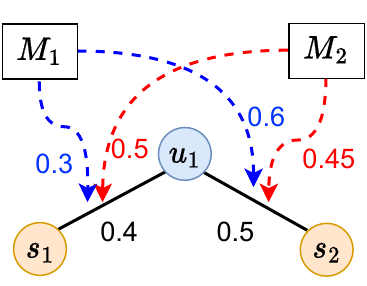}
  \caption{Illustration of the predictions of QoS from two models. The black edge is the response time, which is the QoS. The blue dashed edge represents the predicted response time of $M_1$ model, and the red dashed edge represents the predicted response time of $M_2$ model.}
  \label{fig:example}
\end{figure}

Existing research in the QoS domain predominantly focuses on predicting QoS values. However, high prediction accuracy does not necessarily equate to satisfactory recommendation results. Fig.~\ref{fig:example} depicts a scenario where two users, $u_1$ and $u_2$, interact with two web services, $s_1$ and $s_2$.  user $u_1$ who accesses two web services $s_1$ and $s_2$, yielding observed QoS values for response time of $t_{11}=0.4$ and $t_{12}=0.5$, respectively. Models $M_1$ and $M_2$ predict the QoS grades of $s_1$ and $s_2$ as $\hat{t}_{11}^{M_1}=0.3$, $\hat{t}_{12}^{M_1}=0.6$ and $\hat{t}_{11}^{M_2}=0.5$, $\hat{t}_{12}^{M_2}=0.45$, respectively. Although $M_2$ has better prediction accuracy, recommending $s_j$ to users similar to $u_1$ according to $M_2$ would be inappropriate. This highlights that high prediction accuracy alone does not ensure satisfactory recommendations. Thus, finding similar users or services is essential for better web service recommendations based on QoS.

Collaborative Filtering (CF) has emerged as a key solution to this issue. Leveraging historical user-service interactions, CF provides a more personalized approach to service recommendations. However, the method often encounters the cold-start problem and struggles with data sparsity. To address these limitations, we introduce the use of graph contrastive learning, which has shown promising potential in handling cold-start predictions. 

Graph contrastive learning enables us to learn representations of users and services by contrasting augmented views within a graph structure. By generating diverse perspectives through augmentation, we can better capture the underlying relationships and characteristics of user-service interactions. In our approach, we augment the graph with contextual information, such as geographical locations, to provide a broader and more accurate representation of the interactions. We also use random edge dropping to incorporate randomness into the magnified graph. This additional randomness can account for the complexity of real-world interaction scenarios and simulates the uncertainty and variability present in real-world user-service interactions. This can improve the robustness and adaptability of the model to handle different interaction patterns.

We propose the QoS-aware graph contrastive learning (QAGCL) with geolocation  context for web service recommendations. Our model offers a more effective and comprehensive solution by shifting the focus from QoS prediction to QoS-aware service recommendation, thus improving the quality and accuracy of recommendations.

The main contributions of this paper can be summarized as follows:
\begin{itemize}
    \item We propose the QoS-aware graph contrastive learning (QAGCL) model, a novel approach that combines CF, graph contrastive learning, and contextually augmented graphs.
    \item The QAGCL model effectively mitigates the cold-start problems and data sparsity issues inherent in CF methods.
    \item Our model incorporates contextual information into graph augmentation, which enhances the quality of QoS-based service recommendations.
    \item Through extensive experiments, we demonstrate that our model outperforms several existing models in terms of service recommendation accuracy.
\end{itemize}

\section{Prelimaries \& Related Work}
\subsection{Web Quality of Service (QoS)}
 QoS properties such as response time and throughput have different values depending on the user~\cite{ghafouri2022survey}. When considering a specific service-based application, for example a web service providing video, user-dependent QoS data is mainly determined by the user's calling environment~\cite{zheng2020web}. Therefore, it is generally accepted that if two users have similar past QoS data, they are likely to experience similar QoS in the future due to similar calling environments. From this perspective, a collaborative filtering approach that essentially works by modeling the similarity between users and services~\cite{shin2018LDP,he2019CCA} becomes suitable for QoS prediction\cite{zheng2020web,ghafouri2022survey}. So far, QoS prediction models based on collaborative filtering have made great progress in providing efficient solutions for many service-based applications, such as cloud computing-based applications~\cite{Rehman2015UsersideQF} and multimedia service-based applications~\cite{hossain2014media}.

\subsection{Web Service Recommendation}
In the realm of QoS-based web service recommendation, various studies have explored collaborative filtering techniques to improve recommendation performance~\cite{zheng2012collaborative,chen2015ranking,zheng2020web,yin2019qos,wang2019qos}. Additionally, frameworks utilizing random walks on user-item bipartite graphs have been proposed to predict Web QoS values~\cite{tang2015wswalker,lee2015lmf,ryu2018location}.

However, despite these efforts, data sparsity and the cold-start problem remain significant challenges in web service recommendation~\cite{ryu2018location,zhu2023bgcl}. Data sparsity refers to the scarcity of user-service interaction data, where only a fraction of possible user-service pairs are observed. This sparsity limits the effectiveness of collaborative filtering approaches that heavily rely on historical interactions to make recommendations.

The cold-start problem arises when there is insufficient information about new users or services, making it challenging to provide accurate recommendations. In the context of web service recommendation, this problem can occur when new services are introduced to the system, or when new users join and have limited interaction history.

While existing research has made strides in addressing these challenges, they still pose limitations in terms of handling data sparsity and the cold-start problem effectively. More advanced techniques are required to overcome these obstacles and improve the accuracy and coverage of web service recommendations. Therefore, this study proposes a graph contrastive learning framework for QoS-based web service recommendation. Until recently, QoS values were predicted using graph structures, but there is no example of QoS-based web service recommendation models that apply both graph convolution networks and contrastive learning~\cite{trueman2022graph,chang2021graph}.

\subsection{Graph-based Collaborative Filtering}
Let $\mathbf{R} \in \{0,1\}^{|\mathcal{U}| \times |\mathcal{V}|}$, where $\mathcal{U}$ is a set of users and $\mathcal{V}$ is a set of services, be an interaction matrix. $\mathbf{R}_{u,v}$ is 1 \textit{iff} an interaction $(u,v)$ is observed in data, or otherwise 0. Let $\mathbf{A} \in \mathbb{R}^{N \times N}$ be the adjacency matrix, where $N=|\mathcal{U}|+|\mathcal{V}|$ is the number of nodes. The normalized adjacency matrix is defined as $\tilde{\mathbf{A}} := \mathbf{D}^{-\frac{1}{2}}\mathbf{A}\mathbf{D}^{-\frac{1}{2}}$, where $\mathbf{D} \in \mathbb{R}^{N \times N}$ is the diagonal degree matrix.

The user-service interactions in QoS datasets are closely related to the field of graph-based CF in recommendation systems. We introduce several graph-based CFs, which are mainly used in the general domain of recommender systems suitable for QoS-aware service recommendations. The goal of graph-based CF is to predict user-item (e.g., user-service) ratings by leveraging the relationships and interactions captured in an interaction graph. This approach aims to learn embeddings for users and items in the graph and utilize their inner product to compute the predicted rating.

The user-item relationships can be represented by a bipartite graph and thus, it recently became popular to adopt Graph Convolutional Networks (GCNs) for CF~\cite{Rianne2017GCMC,Wang19NGCF,He20LightGCN,mao2021simplex,Mao21UltraGCN,choi2021ltocf,chen20LRGCCF,kong2022hmlet,liu2021IMP-GCN,choi2023bspm}. GCN is a type of neural network that can operate on graphs. GCN's node embedding is updated by neighbor nodes, and GCN can access $l$-hop neighbor nodes. 

NGCF~\cite{Wang19NGCF} grafted the GCN into CF as it is. LightGCN~\cite{He20LightGCN} has emerged as a standard model by introducing a more lightweight model than NGCF. Its linear GCN layer definition is as follows:
\begin{align}\label{eq:lgc}
    \mathbf{E}^{(l+1)} = \tilde{\mathbf{A}}\mathbf{E}^{(l)},
\end{align} where $\mathbf{E}^{(0)} \in \mathbb{R}^{N \times D}$ is the learnable initial embedding matrix with $D$ dimensions of embedding, and $\mathbf{E}^{(l)}$ denotes the embedding matrix at $l$-th layer.
In the message passing perspective, the GCN layer can be rewritten as follows:
\begin{align}
    \mathbf{e}_i^{(l+1)} = \sum_{j\in\mathcal{N}_i}\frac{1}{d_i d_j}\mathbf{e}_j^{(l)} ,
\end{align}where $\mathbf{e}^{(l)}$ is the feature vector of node $i$ at layer $l$, $\mathcal{N}_i$ is the set of neighbors of node $i$, $d_i$ and $d_j$ are the degrees of nodes $i$ and $j$, respectively. We continue to use matrix-expressed formula of Eq.~\eqref{eq:lgc} later.

The predicted rating, denoted as $\hat{r}$, is calculated by taking the inner product between the user embedding, $\mathbf{e}_u$, and the item embedding, $\mathbf{e}_i$:
\begin{align} 
\hat{r} = \mathbf{e}_u^\textsc{T} \mathbf{e}_i.
\end{align}


\subsection{Contrastive Learning for Recommendation}
In the context of QoS-based web service recommendation, graph contrastive learning is employed to leverage the geolocation context and augment the graph. We introduce several graph contrastive learning methods for recommendation systems. 

Graph-based CFs have demonstrated impressive performance but face challenges such as data sparsity and cold-start problems, as they heavily rely on positive user-item interactions as labels~\cite{yu2018ifbpr,you2020GraphCL}. To overcome these challenges, contrastive learning (CL) for CF methods have been proposed to extract valuable information from unlabeled interactions~\cite{jing2023survey,Wu2021SGL}. These methods utilize different views and contrast them to align node representations, have shown promising results.

SGL~\cite{Wu2021SGL} utilizes LightGCN as its backbone encoder and employs three operators (node dropouts, edge dropouts, and random walks) to generate augmented views. SimGCL~\cite{yu2022SimGCL} simplifies the graph augmentation process by introducing random noises to perturb node representations. LightGCL~\cite{cai2023lightgcl} proposes a graph augmentation strategy based on singular value decomposition to capture global collaborative signals effectively.

These methods consider the views of the same node as positive pairs and views of different nodes as negative pairs. The positive pairs are defined as $\{(\mathbf{e}_{u}', \mathbf{e}_{u}'') | u \in \mathcal{U}\}$, and the negative pairs are defined as $\{(\mathbf{e}_{u}', \mathbf{e}_{v}'') | u, v \in \mathcal{U}, u \neq v \}$. The supervision of positive pairs promotes the similarity between different views of the same user, while the negative pairs encourage the distinction between different nodes. It adopts InfoNCE~\cite{oord2018infonce}, which allows it to learn better user/item representations to preserve node-specific properties and improve generalization ability. The contrastive loss InfoNCE is as follows:
\begin{align}
    \mathcal{L}^{user}_{CL} = \sum_{u\in\mathcal{U}} -\log{\frac{\exp(s(\mathbf{e}_{u}', \mathbf{e}_{u}'')/\tau)}{\sum_{v\in\mathcal{U}}\exp(s(\mathbf{e}_{u}', \mathbf{e}_{v}'')/\tau)}},
\end{align}
where $s(\cdot)$ is the cosine similarity function between two vectors, and $\tau$ is the temperature of softmax, which is a hyper-parameter. It is also applied to items in the same way, which is $\mathcal{L}^{item}_{CL}$. These two loss functions are combined as $\mathcal{L}_{CL} = \mathcal{L}^{user}_{CL} + \mathcal{L}^{item}_{CL}$ and $\mathcal{L}_{CL}$ is the objective function for contrastive learning.

To incorporate geolocation context into the recommendation process, the graph is augmented for contrastive learning techniques. This approach allows for aligning node representations based on different views, leveraging geolocation information alongside other contextual features. By contrasting the augmented views, the model can capture fine-grained similarities and differences between services, enhancing the accuracy and robustness of the recommendation system. Our proposed method utilzes contrastive learning and geolocation-based graph augmentation offers a valuable framework for improving QoS-based web service recommendations.

\section{Motivation}
We describe the motivation for using GCNs, geolocation-based graph augmentation, and contrastive learning for web service recommendations from three perspectives.

\subsection{Why are GCNs suitable for CF?}
Web QoS-based service recommendation heavily relies on understanding the interactions between users and services. GCN excels in capturing the dependencies and relationships within the graph structure, allowing it to model the complex nature of user-service interactions. By leveraging neighborhood information, GCN can effectively propagate and aggregate QoS-related signals across the graph.

In the context of CF, the basic assumption is that \textit{similar users would have similar preferences on items (services)}. GCNs can capture high-order connectivity patterns in the user-item interaction graph and propagate information from neighbors to learn embeddings of users and items. This is especially advantageous in CF since GCNs can capture the similarity between users by propagating information from similar users and learning their embeddings. Thus, incorporating GCNs into CFs offers a natural way to encode collaborative signals in the graph structure of the interaction network.


The integration of GCNs with CF enhances the ability to capture collaborative signals, as it considers not only the user-item interactions but also the inherent structure and connectivity patterns within the graph. This approach effectively leverages high-order connectivity on the graph to better understand user preferences and item characteristics. By incorporating this information, the recommendation system can improve the quality of web service recommendations and deliver more relevant suggestions to users.

\begin{figure*}[t!]
    \centering
    \includegraphics[width=0.95\textwidth]{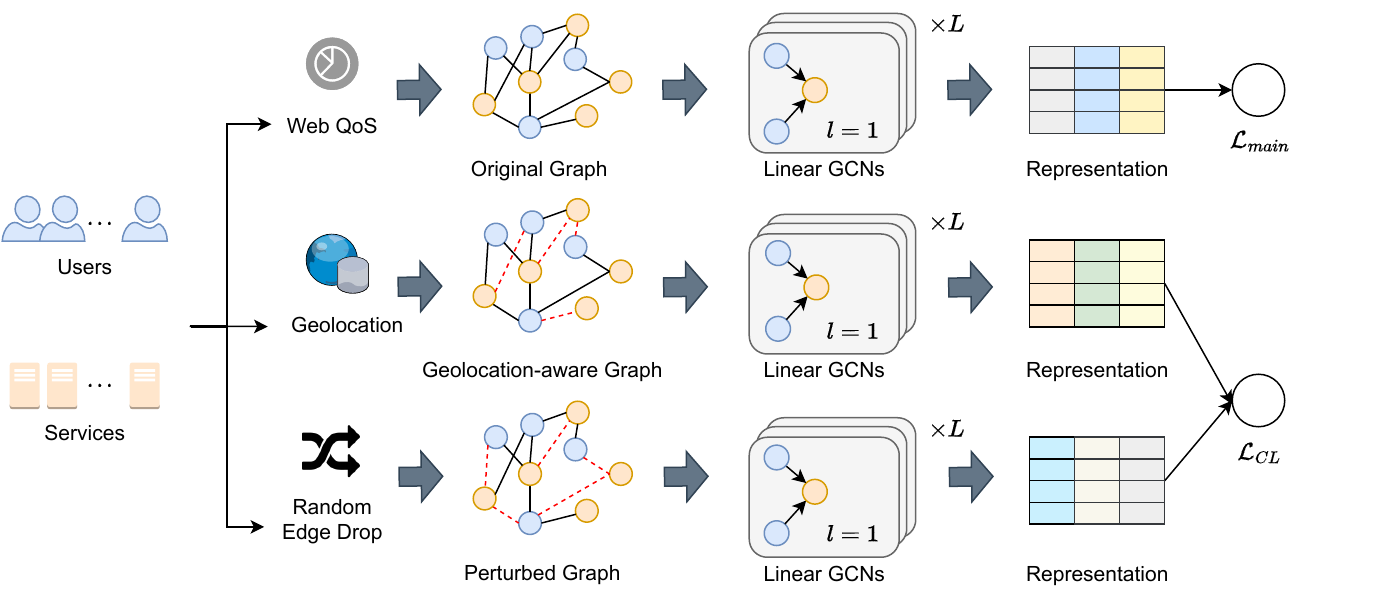}
    \caption{A framework of our proposed QAGCL}
    \label{fig:framework}
\end{figure*}

\subsection{Why is context information such as geolocation important?}
Contextual information is crucial for QoS-based service recommendation due to the following reasons:
\begin{itemize}
    \item Geolocation captures location-specific preferences: Users' service preferences and requirements may vary based on their locations. Incorporating geolocation information enables the recommendation system to consider regional preferences and recommend services that are relevant to users' current or intended locations.
    \item Environmental factors impact service performance: Geolocation information can reflect environmental factors that may affect service performance. For instance, network conditions, availability of resources, and infrastructure quality can vary across different locations. By considering geolocation, the recommendation system can account for these factors and recommend services that are suitable for specific environments.
\end{itemize}

\subsection{Why is it appropriate to perform contrastive learning by creating augmented views in utilizing context information?}
To leverage geolocation information effectively, an augmented view can be constructed by incorporating it into the user-service interaction graph. This augmented view integrates geolocation attributes into the graph representation, creating a richer context-aware representation of the user-service interactions.

Contrastive learning techniques can be applied to the augmented view for QoS-based service recommendation. Contrastive learning aims to learn discriminative representations by contrasting positive (similar) and negative (dissimilar) instances. By using the augmented view, the recommendation system can learn contextual embeddings that capture the geolocation-related patterns and preferences in the user-service interaction graph. Contrastive learning enables the system to effectively model the relationships between services and users in the context of geolocation, enhancing the accuracy of recommendations.

\section{Methododology}
We describe our QAGCL which consists of GCN and a CL framework. We first review its overall architecture and then introduce details.

\subsection{Overall Architecture}
Fig.~\ref{fig:framework} shows our web service recommendation framework named QoS-aware graph contrastive learning (QAGCL). Our overall framework is as follows:
\begin{itemize}
    \item First, we preprocess the user and service invoke data to make a graph structure. We create interactions according to the Web QoS values of users and services. This graph is the original graph used for the recommendation task.
    \item Next, we construct other graph views based on the original graph. We create new graphs based on distance using geolocation information such as latitude and longitude of users and services. We also randomly drop the edges of the graph to create another view of the graph.
    \item Finally, the initial user and service embeddings enter into different GCNs for the three augmented graphs. Each embedding passed through the final layer has a representation of a different view. One of the embeddings from the other view performs the recommendation task, and the other two are used for contrastive learning.
\end{itemize}




\subsection{Graph Augmentation}
We describe the two graph augmentation operators in the following subsubsections.

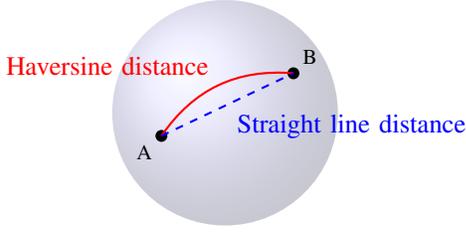
\begin{figure}[t]
  \centering
  \begin{tikzpicture}[scale=1.5]
    \shade[ball color=blue!30, opacity=0.3] (0,0) circle [radius=1cm];
    
    \coordinate (A) at (200:0.6cm);
    \coordinate (B) at (30:0.7cm);
    
    \fill[black] (A) circle [radius=1.5pt] node[below left, font=\footnotesize] {A};
    \fill[black] (B) circle [radius=1.5pt] node[above right, font=\footnotesize] {B};
    

    \draw[red, thick] (A) to[bend left=30] node[midway, above left] {Haversine distance} (B);
    
    
    \draw[dashed, blue, thick] (A) -- (B) node[midway, below right] {Straight line distance};
    
  \end{tikzpicture}
  \caption{Illustration of the Haversine distance and straight line distance between two points on a sphere.}
\end{figure}

\subsubsection{Haversine Distance (HD)}

We use Haversine distance for distance-based data augmentation (See Fig.~\ref{eq:haversine}). For spherical latitude and longitude points, Haversine distance measurements show a high degree of accuracy. Given latitude and longitude coordinates ($a_1$, $b_1$) and ($b_2$, $b_2$), the great circle distance $d$ (km) between the two coordinates can be calculated using the Haversine formula:
\begin{align}\label{eq:haversine}
d &= 2r \arcsin \sqrt{\sin^2\frac{{\Delta_{a}}}{2} + \cos a_1 \cdot \cos a_2 \cdot \sin^2 \frac{{\Delta_{b}}}{2}},
\end{align}
where $r$ is the radius of the Earth (typically taken as 6371 kilometers), $a_1$ and $a_2$ are the latitudes of the two points in radians, $b_1$ and $b_2$ are the longitudes of the two points in radians, and $\Delta_{a}$ and $\Delta_{b}$ mean the difference between $a_1$ and $a_2$, and the difference between $b_1$ and $b_2$, respectively.

We compute the Haversine distance between every user and service. Among the calculated distances, if the distance is greater than a certain threshold $\kappa$, the masking matrix is configured as follows:
\begin{align}
    \mathbf{M}^{(\text{HD})}_{us} = 
\begin{cases}
    1, & \text{if } \frac{d_{us}}{\max(d_{us})}\leq \kappa \\
    0, & \text{otherwise}
\end{cases},
\end{align}where $\mathbf{M}^{(\text{HD})}_{us}$ represents the element of user $u$ and service $s$ of matrix $\mathbf{M}^{(\text{HD})}$. The $\max(d_{us})$ represents the maximum distance value among all distances in the matrix. The threshold $\kappa$ determines the cutoff point for deciding whether the distance is considered large or small. If the ratio of the distance to the maximum distance is less than or equal to the threshold, the element in matrix $\mathbf{M}^{(\text{HD})}$ is set to 1, indicating a small distance. Otherwise, it is set to 0, indicating a large distance.

The augmentation operator for Haversine distance is defined as follows:
\begin{align}
    g_{\text{HD}}(\mathbf{A}) = \mathbf{M}^{(HD)}\odot \mathbf{A},
\end{align}where $\odot$ is a Hadamard product and $\mathbf{M}^{(\text{HD})} \in \{0, 1\}$ is a masking vector on the adjacency matrix. Only partial connections within the neighborhood contribute to the node representations. 

\begin{figure}[t]
    \centering
    \includegraphics[width=0.95\columnwidth]{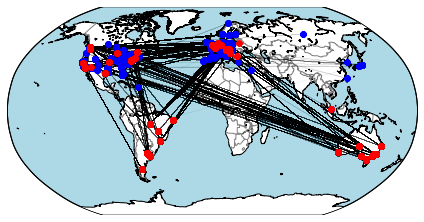}
    \caption{The geolocations information of WSDream. The red (resp. blue) circles are the users' (resp. services') location. The black lines are interactions between users and services.}
    \label{fig:map}
\end{figure}

From a Web QoS perspective, the edge dropout based on Haversine distance offers the following advantages:
\begin{itemize}
    \item Firstly, it allows for realistic modeling by considering the influence of physical proximity on service quality. This aligns with real-world scenarios where distance can impact factors like network latency and bandwidth limitations (See. Fig~\ref{fig:map}).
    \item Second, distance-based edge drop enables localized service recommendations. By focusing services closer to the user, the system can improve the user experience, especially in applications that require low-latency interactions. In fact, as shown in Fig.~\ref{fig:map}, the WSDream dataset needs to be considered for shorter interactions because there are many interactions even when the interaction between the user and the service is far (for example, the distance between continents).
    \item Moreover, incorporating distance in edge drop enhances QoS prediction. By accounting for distance, the recommendation system can better estimate service performance, leading to more accurate predictions and tailored recommendations based on users' geographical context.
\end{itemize}

\subsubsection{Edge Dropout (ED)} It drops out the edges in graph with a dropout ratio $\rho$. The augmentation operator for the edge dropout process is represented as: 
\begin{align}
    g_{\text{ED}}(\mathbf{A}) = \mathbf{M}^{(\text{ED})}\odot \mathbf{A},
\end{align}
where $\mathbf{M}^{(\text{ED})} \in \{0, 1\}$ is the masking vector on the adjacency matrix. Random edge drop is performed in order to simulate noise or uncertainty in the original user-service graph. While the original graph is constructed based on QoS values, it may not account for other contextual information such as distance. By randomly drop edges, we introduce randomness into the graph, which helps to mimic real-world scenarios where the availability or quality of services can vary.

From a Web QoS perspective, random edge drop allows us to model situations where certain services may have intermittent availability or varying performance. In real-world web service environments, factors such as network congestion, server load, or temporary service unavailability can affect the quality of service experienced by users. By randomly dropping edges, we can simulate these scenarios and evaluate the robustness or resilience of recommendation algorithms to such fluctuations in service quality.

Moreover, random edge dropping also helps in assessing the generalization capability of recommendation algorithms. In practice, a recommendation model should be able to provide reasonable recommendations even in the presence of noise or missing information. By introducing random edge drops, we create an environment that tests the ability of the recommendation system to handle incomplete or uncertain data, which is often encountered in real-world web service scenarios.

\subsection{Graph Convolutional Networks}
Mining inherent patterns in graphs is helpful for representation learning. To do this, we use GCNs to capture better user-service interaction, devising GCNs for Haversine distance-based edge drop and random edge drop to create different views for nodes. These GCNs can be represented as follows:
\begin{align}
    \mathbf{E}^{(l)} &= \tilde{\mathbf{A}}\mathbf{E}^{(l-1)},\label{eq:gcn1}\\
    \mathbf{E}'^{(l)} &= \tilde{\mathbf{A}}_{\text{HD}}\mathbf{E}'^{(l-1)},\label{eq:gcn2}\\
    \mathbf{E}''^{(l)} &= \tilde{\mathbf{A}}_{\text{ED}}\mathbf{E}''^{(l-1)},\label{eq:gcn3}
\end{align}where $\mathbf{E}^{(l)}=[\mathbf{E}_u^{(l)}, \mathbf{E}_s^{(l)}]$ is the input feature matrix at layer $l$ and $\tilde{\mathbf{A}}$ is calculated from original bipartite graph $\mathbf{A}$. $\tilde{\mathbf{A}}_{\text{HD}}$ and $\tilde{\mathbf{A}}_{\text{ED}}$ are calculated by augmented operators $g_{\text{HD}}(\cdot)$ and $g_{\text{ED}}(\cdot)$, respectively. Three different views are computed from each of the three different graphs.

\subsection{Prediction Layer}
The prediction layer is the layer to construct the final embedding after propagating through all the $L$ layers. The final embedding uses the weighted sum of the embeddings of each layer:
\begin{align}
    \mathbf{E}_{u}^{(final)} = \sum_{i=0}^{L}w_i\mathbf{E}_{s}^{(i)},\;
    \mathbf{E}_{s}^{(final)} = \sum_{i=0}^{L}w_i\mathbf{E}_{u}^{(i)},\label{eq:wsum1}
\end{align}
where $w_i$ means the weight in each layer. If the $w$ value is the same in all $i$, the average of the embedding values in all layers is used. The weighted sum can achieve good performance by using not only the last layer's embedding, but also the previous layer's embedding. After calculating $\mathbf{E}_{u}^{(final)}$ and $\mathbf{E}_{s}^{(final)}$, the rating of user $u$ for service $s$ is predicted. The dot product of $\mathbf{E}_{u}^{(final)}$ and $\mathbf{E}_{s}^{(final)}$ is performed to predict ratings: 

\begin{align}
    \hat{r}_{u,s} &= \mathbf{e}_{u}^{(final)\textsc{T}} \mathbf{e}_{s}^{(final)}, \label{eq:rating}
\end{align} where $\hat{r}_{u,s}$ is the predicted rating value.

\begin{algorithm}[t]
\SetAlgoLined
\caption{Training algorithm of QAGCL}\label{alg:train}
\KwIn{Web QoS normalized adjacency matrix $\tilde{\mathbf{A}}$, The number of total layers $K$}
Initialize $\mathbf{E}_{u}^{(0)}$ and $\mathbf{E}_{s}^{(0)}$;\\
Generate augmented graphs $\tilde{\mathbf{A}}_{\text{HD}}$ and $\tilde{\mathbf{A}}_{\text{ED}}$;\\
\While {the joint loss $\mathcal{L}$ is not converged}{
    \For{$l \leftarrow 1 \; \text{to} \; L$}{ 
        $\mathbf{E}_{u}^{(l)}, \mathbf{E}_{s}^{(l)} =  \text{Eq}$.~\eqref{eq:gcn1}, $\text{Eq}$.~\eqref{eq:gcn1} with $\tilde{\mathbf{A}}$\;
        $\mathbf{E}^{'(l)}_{u}, \mathbf{E}^{'(l)}_{s} =  \text{Eq}$.~\eqref{eq:gcn2}, $\text{Eq}$.~\eqref{eq:gcn2} with $\tilde{\mathbf{A}}_{\text{HD}}$\;
        $\mathbf{E}^{''(l)}_{u}, \mathbf{E}^{''(l)}_{s} =  \text{Eq}$.~\eqref{eq:gcn3}, $\text{Eq}$.~\eqref{eq:gcn3} with $\tilde{\mathbf{A}}_{\text{ED}}$\;
    }
    $\mathbf{E}_{u}^{(final)}, \mathbf{E}_{s}^{(final)} = \text{Eq}.~\eqref{eq:wsum1},\text{Eq}.~\eqref{eq:wsum1}$\;
    $\mathbf{E}_{u}^{'(final)}, \mathbf{E}_{s}^{'(final)} = \text{Eq}.~\eqref{eq:wsum2},\text{Eq}.~\eqref{eq:wsum2}$\;
    $\mathbf{E}_{u}^{''(final)}, \mathbf{E}_{s}^{''(final)} = \text{Eq}.~\eqref{eq:wsum3},\text{Eq}.~\eqref{eq:wsum3}$\;
    $\hat{r}_{u,s} = \text{Eq}$.~\eqref{eq:rating}\;
    Compute the joint objective loss with\;
    Update $\mathbf{E}_u^{(0)}$ and $\mathbf{E}_s^{(0)}$ with joint loss\;
}
\Return $\mathbf{E}_u^{(0)}$ and $\mathbf{E}_s^{(0)}$;
\end{algorithm}

\subsection{Contrastive Learning}
We use two generated views from Eqs.~\eqref{eq:gcn2} and~\eqref{eq:gcn3} as follows:
\begin{align}
    \mathbf{E}_{u}^{'(final)} = \sum_{i=0}^{L}w_i\mathbf{E}_{s}^{'(i)},\;
    \mathbf{E}_{s}^{'(final)} = \sum_{i=0}^{L}w_i\mathbf{E}_{u}^{'(i)},\label{eq:wsum2}\\
    \mathbf{E}_{u}^{''(final)} = \sum_{i=0}^{L}w_i\mathbf{E}_{s}^{''(i)},\;
    \mathbf{E}_{s}^{''(final)} = \sum_{i=0}^{L}w_i\mathbf{E}_{u}^{''(i)}.\label{eq:wsum3}
\end{align}

After generating the different two views, we employ a contrastive objective that enforces the filtered representations of each node in the two views to agree with each other. We perform the CL training by directly contrasting the distance-based augmented view  $\mathbf{e}'^{(final)}$  with the random edge drop-based augmented view  $\mathbf{e}''^{(final)}$  using the InfoNCE~\cite{oord2018infonce} loss: 
\begin{align}\label{eq:contrastive_loss}
\mathcal{L}_{CL} = \sum_{i \in \mathcal{B}} -log\frac{\exp(\text{sim}(\mathbf{e}'^{(final)}_{i},\mathbf{e}''^{(final)}_{i})/\tau)}{\sum_{j \in \mathcal{B}}\exp(\text{sim}(\mathbf{e}'^{(final)}_{i},\mathbf{e}''^{(final)}_{j}/\tau)},
\end{align}where $i$, $j$ are a user and an item in a sampled batch $\mathcal{B}$, and $\mathbf{e}'^{(final)}_{i}$,  $\mathbf{e}''^{(final)}_{i}$, and $\mathbf{e}''^{(final)}_{j}$ are node representations from Eqs.~\eqref{eq:wsum2} and~\eqref{eq:wsum3}.

\subsection{How to Train}


We use the Bayesian Personalized Ranking (BPR) loss function~\cite{rendle2009BPR} together with Eq.~\eqref{eq:contrastive_loss}.
As shown in Eq.~\eqref{eq:obejective_function}, therefore, our joint learning objective is as follows:
\begin{align}\label{eq:obejective_function}
\mathcal{L}=\mathcal{L}_{main}+\lambda_1 \cdot \mathcal{L}_{CL}+\lambda_2 \cdot \Vert\mathbf{\Theta}\Vert^2_2,
\end{align}which consists of the Bayesian personalized ranking (BPR) loss $\mathcal{L}_{main}$ and the CL loss $\mathcal{L}_{CL}$. The hyperparameters $\lambda_1$ and $\lambda_2$ control the trade-off among the two loss functions and the regularization term. $\mathbf{\Theta}$ denotes the embeddings to learn, i.e, $\mathbf{\Theta} = \mathbf{E}(0)$ in our framework. $\mathcal{L}_{main}$ is defined as:
\begin{align}
\mathcal{L}_{main} = - \sum_{(u,i,j) \in \mathcal{B}} \log (\sigma (\hat{r}_{ui} - \hat{r}_{uj})),
\end{align}where $\sigma$ is the sigmoid function, $\hat{r}_{ui}$ and $\hat{r}_{uj}$ denote the predicted rating scores for a pair of positive and negative services of user $u$.
 
After minimizing the joint loss in Eq.~\eqref{eq:obejective_function}, we use the output embeddings of the GCN, i.e., $\mathbf{E}_{u}^{(final)}$ and $\mathbf{E}_{i}^{(final)}$, as the final representation. The exact training method is described in Alg.~\ref{alg:train}.

\section{Experiments}
To justify the superiority of QAGCL and reveal the reasons of its effectiveness, we conduct extensive experiments and answer the following research questions:
\begin{enumerate}
    \item \textbf{RQ1:} How does QAGCL perform w.r.t. top-K recommendation as compared with the CF models?
    \item \textbf{RQ2:} How effectively does the QAGCL model mitigate the cold-start problem in comparison to the existing methodologies?
    \item \textbf{RQ3:} How does the QAGCL model affect the quality of QoS-based service recommendations based on graph augmentation techniques?
    \item \textbf{RQ4:} How does varying the number of graph layers affect the performance of the proposed QAGCL?
\end{enumerate}

\subsection{Experimental Settings}
\subsubsection{Datasets}
The web service QoS dataset used in the experiment is WSDream~\cite{zheng2012collaborative}, and response time is used as the QoS value\footnote{\url{http://wsdream.github.io/dataset} The dataset is available for download.}. The dataset used in the experiment is the same as Table~\ref{tab:dataset}, and the test set ratio is 20\%. In order to construct the dataset, it is assumed that there is connectivity when response time $t_{ij}$ is lower than a certain threshold. For example, if $t_{ij}$ is less than $\gamma$, it is regarded as a positive interaction and a graph is constructed. We set the threshold $\gamma=0.05s$. While the number of interactions in the warm-start environment is 57,727, there are two datasets for the cold-start environment, each with 8,490 and 1,036 interaction matrices. Cold-start-ex means that the density is 2.75\%, which is more extreme than Cold-start, which is 5.36\%. In the case of the warm-start environment, one user configured more than 10 interactions so that training and test datasets can be configured. In the case of cold-start environment, filtering was performed so that there were more than 2 interactions.

\subsubsection{Evaluation Metrics}
For each user in the test set, all non-interaction services are considered as negative samples, and the model calculates the user's rating for all samples except for the positive samples used in the training dataset. 
To compare the performance of our model with the baseline model for top-$K$ recommendations, we use Recall@$K$ and NDCG@$K$, commonly used in rank-based evaluation, as evaluation metrics. Recall@$K$ represents the ratio of $K$ recommended services out of all services and is defined as follows:
\begin{align}
    \text{Recall@}K=\frac{\text{rel}_K}{\min(K,\text{rel})}.
\end{align} $\text{rel}_K$ means the number of related items in the top-$K$ results, and $\text{rel}$ means the total number of related items for the user. Recall@$K$ means the ratio of how many $K$ systems recommended by the model are included among all services associated with the user.
Normalized Discounted Cumulative Gain (NDCG) evaluates the difference between a list of recommended items and a list of optimally ranked items and is defined as:
\begin{align}
    \text{NDCG@}K=\frac{\text{DCG@}K}{\text{IDCG@}K}.
\end{align} NDCG@$K$ evaluates performance by weighting the order of recommendations, and the closer to 1, the better the performance. DCG@$K$ and IDCG@$K$ are DCG (Discounted Cumulative Gain) of the top $k$ items of predicted rank and ideal rank, respectively. DCG@$K$ is calculated as:
\begin{align}
    \text{DCG@}K=\sum_{i=1}^{K}\frac{2^{\text{rel}_{i} - 1}}{\log_2(i+1)},
\end{align}where $\text{rel}_i$ is the value of the item at the rank position. The value of NDCG is between 0 and 1, and the higher the value, the higher the rank, and 1 represents the ideal rank.

\begin{table}[t]
    \centering
    \caption{Dataset information of WSDream}
    \begin{tabular}{c ccc}
        \toprule
        \multirow{2}{*}{Dataset}   &  WSDream & WSDream & WSDream\\
                 &  (Warm-start) & (Cold-start) & (Cold-start-ex)\\ \midrule
        \# Users & 338 & 275 & 172 \\ 
        \# Services & 5,824 & 575 & 219\\ 
        $\gamma$ & 0.05 & 0.02 & 0.01\\
        Core & 10 & 2 & 2\\
        Density & 13.64\% & 5.36\% & 2.75\%\\
        \# Interactions & 57,727 & 8,490 & 1,036\\
        \bottomrule
    \end{tabular}
    \label{tab:dataset}
\end{table}

\begin{table*}[t]
    \centering
    \caption{Comparison of Recall@K and NDCG@K recommendation performance of each model for WSDream (Warm-start). \textbf{boldface} is the best performance, \underline{underline} is the second-best performance. \textit{Improvement} stands for the improvement over second-best recommendation performance.}
    \begin{tabular}{c cc cc}\toprule
        \multirow{2}{*}{Model}  & \multicolumn{4}{c}{WSDream (Warm-start)} \\ \cmidrule(lr){2-3}\cmidrule(lr){4-5}
                                & Recall@20 & NDCG@20 & Recall@40 & NDCG@40\\ \midrule
        UMEAN                   & 0.0943    & 0.2134
                                & 0.0989    & 0.2193\\
        IMEAN                   & 0.0884    & 0.2009
                                & 0.0919    & 0.2074\\ \midrule
        BPR-MF                  & 0.2012    & 0.4170
                                & 0.3390    & 0.4483\\
        NeuMF                   & 0.1950    & 0.4104
                                & 0.3041    & 0.4198\\ \midrule
        NGCF                    & 0.2095    & 0.4294
                                & 0.3409    & 0.4511  \\
        LightGCN                & 0.2113    & 0.4325
                                & 0.3419    & 0.4595  \\
        \midrule
        SGL                     & \underline{0.2158}    & \underline{0.4628}
                                & \underline{0.3569}    & \underline{0.4958}  \\
        SimGCL                  & 0.2150    & 0.4563
                                & 0.3566    & 0.4910  \\
        LightGCL                & 0.1946    & 0.4268
                                & 0.2974    & 0.4271  \\\midrule
        QAGCL                   & \textbf{0.2212}    & \textbf{0.4751}
                                & \textbf{0.3825}    & \textbf{0.5149}  \\
        \midrule
        \textit{Improvement} & 2.50\% & 2.66\%
                             & 7.17\% & 3.85\%\\
        \bottomrule
    \end{tabular}
    \label{tbl:main_exp}
\end{table*}

\begin{table*}[t]
    \centering
    \caption{Comparison of Recall@K and NDCG@K recommendation performance of each model for WSDream (Cold-start)}
    \begin{tabular}{c cc cc}\toprule
        \multirow{2}{*}{Model}  & \multicolumn{4}{c}{WSDream (Cold-start)} \\ \cmidrule(lr){2-3}\cmidrule(lr){4-5}
                                & Recall@20 & NDCG@20 & Recall@40 & NDCG@40\\ \midrule
        UMEAN                   & 0.2193    & 0.1843
                                & 0.2443    & 0.2022\\
        IMEAN                   & 0.2034    & 0.1792
                                & 0.2431    & 0.1984\\ \midrule
        BPR-MF                  & 0.4376    & 0.3767
                                & 0.6321    & 0.4363\\
        NeuMF                   & 0.4012    & 0.3551
                                & 0.6104    & 0.4111\\ \midrule
        NGCF                    & 0.5015    & 0.4532
                                & 0.6459    & 0.4913  \\
        LightGCN                & 0.5751    & 0.5009
                                & 0.7274    & 0.5513  \\
                                \midrule
        SGL                     & 0.6123    & \underline{0.5717}
                                & 0.7897    & \underline{0.6251}  \\
        SimGCL                  & \underline{0.6388}    & 0.5702
                                & \underline{0.8002}    & 0.6249  \\
        LightGCL                & 0.6077    & 0.4985
                                & 0.7516    & 0.5466  \\\midrule
        QAGCL                   & \textbf{0.6426}    & \textbf{0.5845} 
                                & \textbf{0.8300}    & \textbf{0.6450}\\
        \midrule
        \textit{Improvement} & 0.59\% & 2.24\%
                             & 3.72\% & 3.18\%\\
        \bottomrule
    \end{tabular}
    \label{tbl:main_exp_cold}
\end{table*}

\begin{table*}[t]
    \centering
    \caption{Comparison of Recall@K and NDCG@K recommendation performance of each model for WSDream (Cold-start-ex)}
    \begin{tabular}{c cc cc}\toprule
        \multirow{2}{*}{Model}  & \multicolumn{4}{c}{WSDream (Cold-start-ex)} \\ \cmidrule(lr){2-3}\cmidrule(lr){4-5}
                                & Recall@20 & NDCG@20 & Recall@40 & NDCG@40\\ \midrule
        UMEAN                   & 0.2012    & 0.1204
                                & 0.2444    & 0.1455\\
        IMEAN                   & 0.1994    & 0.1195
                                & 0.2402    & 0.1412\\ \midrule
        BPR-MF                  & 0.4284    & 0.2585
                                & 0.5800    & 0.2999\\
        NeuMF                   & 0.4020    & 0.2498
                                & 0.5712    & 0.2901\\ \midrule
        NGCF                    & 0.7412    & 0.5066
                                & 0.7865    & 0.5150  \\
        LightGCN                & 0.9026    & \underline{0.7204}
                                & 0.9404    & \underline{0.7337}  \\
                                \midrule
        SGL                     & 0.8997    & 0.7155
                                & \underline{0.9471}    & 0.7303  \\
        SimGCL                  & \underline{0.9139}    & 0.7185
                                & 0.9410    & 0.7291  \\
        LightGCL                & 0.8083    & 0.6493
                                & 0.8940    & 0.6749  \\\midrule
        QAGCL                   & \textbf{0.9178}    & \textbf{0.7430}
                                & \textbf{0.9491}    & \textbf{0.7554}  \\
        \midrule
        \textit{Improvement} & 0.42\% & 3.14\%
                             & 0.21\% & 2.96\%\\
        \bottomrule
    \end{tabular}
    \label{tbl:main_exp_cold-ex}
\end{table*}

\subsubsection{Compared Baselines}
We compare our model against 9 baselines with different learning paradigms:
 \begin{itemize}
    \item \textbf{Traditional baseline}: UMEAN predicts missing values by averaging the available QoS values based on the target user. IMEAN predicts missing values by averaging the available QoS values based on the target Web service.
    \item \textbf{Matrix factorization}: BPR-MF~\cite{rendle2009BPR} is a classical collaborative filtering algorithm that minimizes a pair-wise loss function to learn implicit feedback. In BPR, MF is used to initialize the embedding of users and items. NeuMF~\cite{He2017NCF} is a collaborative filtering algorithm that uses non-linear hidden layers on the interaction of users and item embeddings to capture interactions. 
    \item \textbf{Graph-based collaborative filtering}: NGCF~\cite{Wang19NGCF} and LightGCN~\cite{He20LightGCN}.
     \item \textbf{Contrastive Learning for Collaborative Fitlering}: SGL~\cite{Wu2021SGL}, SimGCL~\cite{yu2022SimGCL}, and LightGCL~\cite{cai2023lightgcl}.
 \end{itemize}

\subsubsection{Hyperparameters}
For fair comparison with previous studies, we set the size of embedding $D$ as 64, the number of epochs as 100, and the same test split ratio as 0.2. We also further search the best hyperparameters for baselines based on their recommended settings. For our method, we test the following hyperparameters:
\begin{itemize}
    \item The learning rate is in \{\num{1.0e-4}, \num{5.0e-4}, \num{1.0e-3}, \num{5.0e-3}, \num{1.0e-2}\};
    \item The regularization weight for the InfoNCE loss $\lambda_1$ is in $\{0.1, 0.2,\cdots, 1.0\}$;
    \item The regularization weight $\lambda_2$ is in $\{\num{1.0e-7},\num{1.0e-6},\num{1.0e-5}\}$.
    \item The ratio $\rho$ for $g_{\text{ED}}$ is in $\{0.1, 0.2,\cdots, 0.5\}$;
    \item The ratio $\kappa$ for $g_{\text{ED}}$ is in $\{0.1, 0.2,\cdots, 0.5\}$;
    \item The number of layer $L$ is in $\{1,2,3,4\}$;
\end{itemize}


\subsection{Evaluation of Top-K Recommendation Performance (\textbf{RQ1})}
In the WSDream (Warm-start) dataset of Table~\ref{tbl:main_exp}, we compare the recommendation performances between our proposed QAGCL method and baseline models.  QAGCL shows the best performance in all evaluation metrics. Based on Recall@40, QAGCL improved by 7.17\% over SGL, and based on NDCG@40, it showed a performance improvement of 3.85\% over the recommended performance of BPR-MF. Among the baseline models, recommendation models based on graph contrastive learning generally show better results. This shows that the effect of contrastive learning works in performing web service recommendation on the user-service interaction dataset.

In contrast, graph convolution-based CFs show superior recommendation performance compared to BPR-MF and NeuMF. This difference can be attributed to the limitations of the MF-based methods in effectively utilizing neighbor information and high-order connectivity of users and services. NGCF and LightGCN, on the other hand, exhibit slightly improved performance due to their explicit exploration of higher-order connections within the neighborhood. UMEAN and IMEAN notably show relatively lower recommendation performance than the other baselines.

\subsection{Cold-Start Problem Mitigation (\textbf{RQ2})}
We test the effectiveness of the QAGCL model and baseline models in mitigating the cold-start problem by configuring the cold-start environment by tuning the sparsity of WSDream dataset. In Tables~\ref{tbl:main_exp_cold} and~\ref{tbl:main_exp_cold-ex}, we compare the results of the WSDream (Cold-start) dataset, the cold-start setting of WSDream, and the extreme cold-start setting, WSDream (Cold-start-ex).

\subsubsection{Result of WSDream (cold-start)}
In Table~\ref{tbl:main_exp_cold}, QAGCL shows the best performance in all baselines. Based on Recall@40, the cold-start dataset showed an improvement of 3.72\% compared to SimGCL. In this cold-start environment, contrastive learning-based models show better performance than graph convolution-based CF methods.

\subsubsection{Result of WSDream (Cold-start-ex)}
Table ~\ref{tbl:main_exp_cold-ex} also shows that QAGCL shows the highest recommendation performance in all evaluation scales, improving by 3.14\% over LightGCN and 3.41\% over SimGCL based on NDCG@20. In addition, LightGCN shows the second-best Recall@40 performance of 0.7337, while QAGCL shows a recommendation performance of 0.7554, an improvement of 2.96\%.

Through this experiment, our proposed QAGCL shows a better method even in a cold-start environments with sparse interactions in the dataset for \textbf{RQ2}. Therefore, the necessity of our proposed design can be confirmed in both warm-start and cold-start environments.

\subsection{Impact of Graph Augmentation Techniques on QoS-Based Service Recommendations (\textbf{RQ3})}

As shown in Table~\ref{tab:aug}, using a combination of HD and ED augmetnation operators for QoS-based service recommendations is effective. The HD takes into account the physical proximity of services. The ED reflects the unpredictable nature of service usage patterns and incorporates diverse perspectives. The randomness of ED allows the QAGCL to adapt to different user behaviors and environmental factors, enhancing the robustness and generalization ability of the recommendation system. By combining these approaches, the augmented graph benefits from localized proximity information and adaptability to diverse user preferences and network conditions. This hybrid approach enhances recommendation performance from a Web QoS perspective.

\begin{table}[t]
    \centering
    \caption{Graph augmentation technique comparison experiments on WSDream (Cold-start). ND stands for a random node drop.}
    \begin{tabular}{ccc}\toprule
        Graph Augmentation & Recall@20 & NDCG@20 \\\midrule
        HD \& ED & 0.6426 & 0.5845\\
        HD \& ND & 0.5958 & 0.4977\\
        ED \& ED & 0.6105 & 0.5698\\
        \bottomrule
    \end{tabular}
    \label{tab:aug}
\end{table}

\subsection{Sensitivity Study on the Number of Graph Layers (\textbf{RQ4})}
To understand how the different setting influenced the effectiveness of the QAGCL, we varied the number of graph layers of our methods. When the number of graph convolution layers $L$ is 2, the best performance is shown for Recall@40 and NDCG@40. However, in the case of Recall@20, it is best when $L=2$. And when $L=4$, recommendation performance tends to drop. Through this, the best performance is shown with an optimal $L$.


\begin{table}[t]
    \centering
    \caption{Sensitivity analysis of the number of graph layers for WSDreadm (Cold-start)}
    \begin{tabular}{c cccc}\toprule
        $L$ & Recall@20 & NDCG@20 & Recall@40 & NDCG@40 \\\midrule
        1 & 0.5394 & 0.4718 & 0.7223 & 0.5303\\
        2 & 0.6524 & 0.5783 & 0.8109 & 0.6318\\
        3 & 0.6426 & 0.5845 & 0.8300 & 0.6450\\
        4 & 0.6342 & 0.5756 & 0.8274 & 0.6404\\
        \bottomrule
    \end{tabular}
    \label{tab:layer}
\end{table}

\section{Threads to Validity}

The threat to construction validity lies in the data preprocessing step to construct the bipartite graph. For graph construction, it is assumed that there is an interaction between a user and a service with a lower response time based on the threshold $\gamma$. Classifying whether or not there is an interaction as 1 or 0 may limit the rich use of web service QoS data. To overcome this limitation, we will design the method using bipartite graphs with weights in the future.

The threat to internal validity is not using other geolocation-based recommendation system techniques as a comparison model. However, the goal of our study is to explore how contrastive learning by augmenting the graph is an effective design for web service recommendation. Thus, we focused on comparing with contrastive learning techniques, and plan to conduct additional experiments and compare various distance-based models in the future.


\section{Conclusion and Future Work}
In this paper, we proposed the QoS-aware graph contrastive learning (QAGCL) model for web service recommendation. Our model addressed the limitations of CF methods by incorporating graph contrastive learning, contextual augmentation, and random edge dropout. Through extensive experiments, we demonstrated the effectiveness of the QAGCL model in improving web service recommendation accuracy. By leveraging graph contrastive learning, our model effectively handled the cold-start problem and data sparsity issues, providing more accurate recommendations. The incorporation of contextual augmentation, including geographical information, allowed for a broader perspective of user-service interactions. The results of our experiments showed that the QAGCL model outperformed several existing models in terms of recommendation accuracy.

Future work could explore further enhancements to the QAGCL model, such as incorporating additional contextual information or exploring different techniques for graph augmentation and contrastive learning. 

\bibliographystyle{ieeetr}
\bibliography{ref}

\end{document}